\documentclass[conference, 10pt, letterpaper]{IEEEtran}

\usepackage{amsmath}
\usepackage{amsfonts}
\usepackage{algorithmic}
\usepackage{graphicx}
\usepackage{textcomp}
\usepackage{xcolor}
\usepackage{acronym}
\usepackage{subfig}
\usepackage{hyperref}

\usepackage{listings}

\usepackage{graphicx}
\graphicspath{ {./images/} }

\colorlet{punct}{red!60!black}
\definecolor{background}{HTML}{EEEEEE}
\definecolor{delim}{RGB}{20,105,176}
\colorlet{numb}{magenta!60!black}

\lstdefinelanguage{json}{
    basicstyle=\normalfont\ttfamily,
    numbers=left,
    numberstyle=\scriptsize,
    stepnumber=1,
    numbersep=8pt,
    showstringspaces=false,
    breaklines=true,
    frame=lines,
    backgroundcolor=\color{background},
    literate=
     *{:}{{{\color{punct}{:}}}}{1}
      {,}{{{\color{punct}{,}}}}{1}
      {\{}{{{\color{delim}{\{}}}}{1}
      {\}}{{{\color{delim}{\}}}}}{1}
      {[}{{{\color{delim}{[}}}}{1}
      {]}{{{\color{delim}{]}}}}{1},
}

\usepackage{todonotes}

\usepackage{tikz}
\usetikzlibrary{arrows}

\acrodef{PPFS}[DLPFS]{Data Leakage Prevention FileSystem}
\acrodef{GDPR}{General Data Protection Regulation}
\acrodef{FUSE}{File system in User Space}
\acrodef{JSON}{JavaScript Object Notation}
\acrodef{HDB}{Hippocratic databases}
\acrodef{KB}{Knowledge Base}
\acrodef{ICD}{International Classification of Diseases}
\acrodef{WHO}{World Health Organization}
\acrodef{VM}{Virtual Machine}
\acrodef{NFS}{Network FileSystem}
\acrodef{CSV}{Comma-Separated Values}
\acrodef{OS}{Operating System}
\acrodef{SAN}{Storage Area Network}
\acrodef{HDFS}{Hadoop Distributed FileSystem}
\acrodef{LB}{LoopBack}
\acrodef{DLP}{Data Leakage Prevention}

\begin{document}

\title{DLPFS: The Data Leakage Prevention FileSystem}


\author{\IEEEauthorblockN{Stefano Braghin}
\IEEEauthorblockA{IBM Research Europe\\Dublin -- Ireland}
\and
\IEEEauthorblockN{Marco Simioni}
\IEEEauthorblockA{IBM Research Europe\\Dublin -- Ireland}
\and
\IEEEauthorblockN{Mathieu Sinn}
\IEEEauthorblockA{IBM Research Europe\\Dublin -- Ireland}}


\maketitle

\begin{abstract}
Shared folders are still a common practice for granting third parties access to data files, regardless of the advances in data sharing technologies.
Services like Google Drive, Dropbox, Box, and others, provide infrastructures and interfaces to manage file sharing.
%
The human factor is the weakest link and data leaks caused by human error are regrettable common news.
This takes place as both mishandled data, for example stored to the wrong directory, or via misconfigured or failing applications dumping data incorrectly.
We present \ac{PPFS}, a first attempt to systematically protect against data leakage caused by misconfigured application or human error.
This filesystem interface provides a privacy protection layer on top of the POSIX filesystem interface, allowing for seamless integration with existing infrastructures and applications, simply augmenting existing security controls.
At the same time, \ac{PPFS} allows data administrators to protect files shared within an organisation by preventing unauthorised parties to access potentially sensitive content.
\ac{PPFS} achieves this by transparently integrating with existing access control mechanisms.
We empirically evaluate the impact of \ac{PPFS} on system's performances to demonstrate the feasibility of the proposed solution.
\end{abstract}

\section{Introduction}
\label{sec:intro}
Most of today's data breaches are due to human error caused by insiders (e.g. misconfiguration, poor data governance), rather than attacks by hackers from outside an organization~\cite{report_misconfiguration}.

Incorrectly configured applications and bugs are an ever present threat to confidentiality of sensitive data.
Examples of these scenarios include log files, which might contain incorrectly handled log level messages, and thus potentially leaking sensitive information such as usernames and passwords, and stack traces or core dumps of crashed applications.

Several approaches have been taken to address the issue, mainly through access control or encryption, hence by restricting who can access specific storage structures (e.g.~partitions, mount points, directories, files and/or zones).
This is still not sufficient if data are meant to be shared among various principals in a platform, in case of data that is required to be accessible for various reasons (e.g.~log files that need to be accessible both for audit and debugging purposes), or if datasets is incorrectly stored in locations not initially envisioned -- like public nodes of a Hadoop cluster in a hybrid cloud setting.

Several common use cases require data, possibly containing sensitive information, to be accessible from different user/roles with different granularity/level of completeness.

Currently, the solution most used in practice is to create different versions of the dataset for each purpose, which is expensive or even impractical if large volumes of data need to be replicated.
Alternatives, such as utilization of techniques based on fully or partially homomorphic encryption have been proposed~\cite{10.1145/2347673.2347681}.
These solutions incur significant performance penalties, however, caused by the mathematical complexity of the algorithms required to achieve required levels of security.

As the adoption of tools like Dropbox\footnote{\url{https://www.dropbox.com}}, Google Drive\footnote{\url{https://www.google.com/drive}}, and Box\footnote{\url{https://www.box.com/}} suggests, 
file systems offer a very popular approach to data sharing across applications~\cite{10.1145/3126908.3126924} and across systems~\cite{allison1999file}.

Therefore, we propose \ac{PPFS}, a novel mechanism to share data across multiple applications and systems leveraging state-of-the-art data type identification and de-identification technologies.
\ac{PPFS} exposes a POSIX file system API to applications accessing a protected subtree of the file system.
Practically, \ac{PPFS} acts as a middleware between applications and the actual file systems, identifying and protecting sensitive data on both read and write paths.

\ac{PPFS} allows data users to share data in a privacy preserving fashion across multiple systems without the need to create bespoke copies of the data for the target application.
Moreover, \ac{PPFS} allows legacy applications to operate on data de-identified on the fly, without the need of modifying such applications themselves.
This removes the burden of modifying legacy and mission critical applications from the developers, allowing DevOps and SecOps teams to define fine grained access control and privacy profiles, according to application and context requirements.

The rest of the paper is organized as follows.
Section~\ref{sec:method} introduces the design principles of \ac{PPFS} and its operational steps.
Section~\ref{sec:implementation} and~\ref{sec:evaluation} present the implementation of the \ac{PPFS} prototype and discuss empirical performance evaluations.
Finally, Section~\ref{sec:related} compares our solution with the state of the art and Section~\ref{sec:conclusion} summarizes the contribution and depicts possible future directions.

\section{\acl{PPFS} in Practice}
\label{sec:method}
\begin{figure}[tb]
    \centering
    \includegraphics[width=.8\columnwidth]{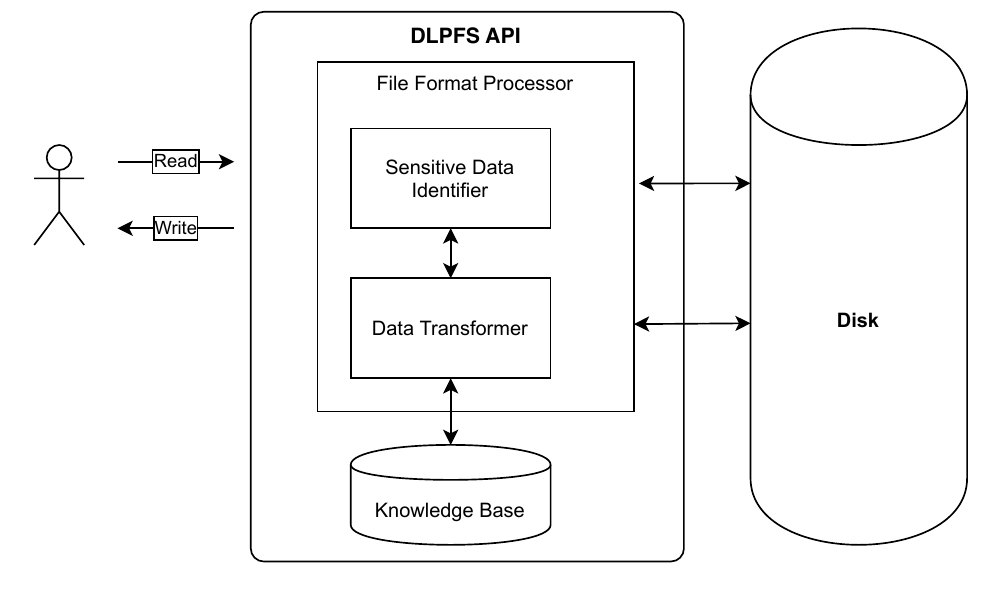}
    \caption{Overall modules architecture}
    \label{fig:architecture}
\end{figure}

\ac{PPFS} operates as a middleware between a software application and the file system stack.
The \ac{PPFS} conceptual architecture is shown in Figure~\ref{fig:architecture}.

\begin{figure}[t]
    \centering
    \includegraphics[width=\columnwidth]{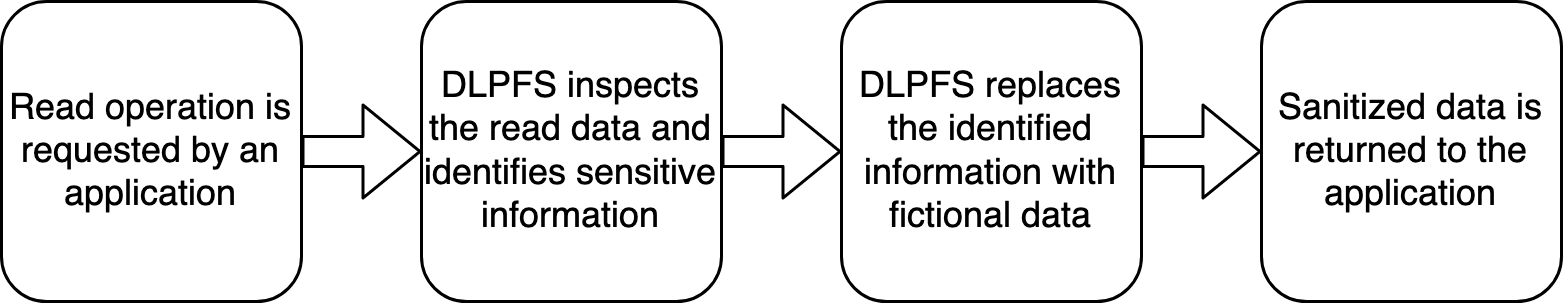}
    \caption{Read data flow}
    \label{fig:read}
\end{figure}

\begin{figure}[t]
    \centering
    \includegraphics[width=\columnwidth]{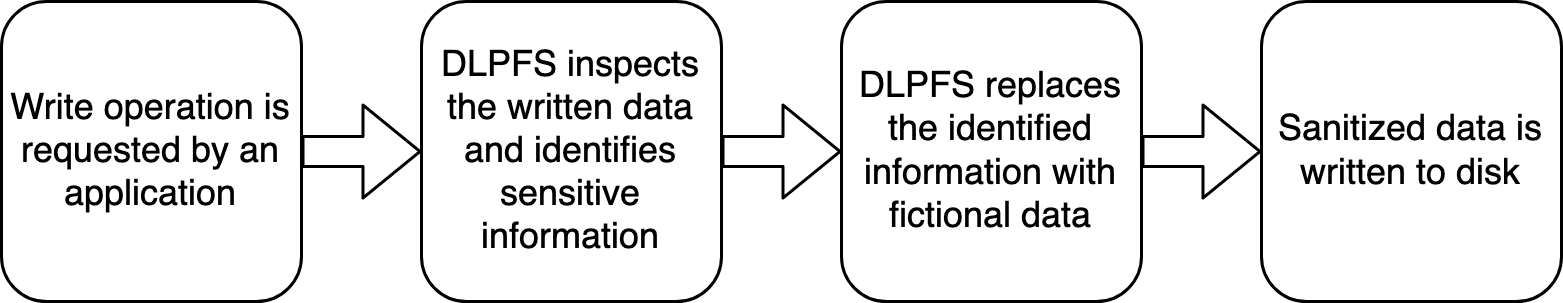}
    \caption{Write data flow}
    \label{fig:write}
\end{figure}

The system exposes a POSIX filesystem API that proxies all requests performed by a client application to the actual filesystem.
\ac{PPFS} intercepts all read and write operations and acts on the transferred data according to the instructions specified in the knowledge base.
%
%
The two supported read and write flows are sketched in Figure~\ref{fig:read} and Figure~\ref{fig:write}.

The key idea behind \ac{PPFS} is to intercept and analyse data as it is transferred between the data storage and applications.
Inspection and transformation operations can generally be applied to streams of raw data.
Hence, there is no strict need for \ac{PPFS} to be aware of the format of the files where the data is being read from or written to, or their structure.

However, having knowledge of the file structure improves how information is handled in specific scenarios, leading to more accurate data detection and data transformation. As an example, consider an application that loads data from a \ac{CSV} file into memory by sequentially reading blocks of $1,024$ bytes of data. According to the \ac{CSV} format, information is stored within data fields that are separated by a comma character (`\verb$,$') and groups of fields, i.e. rows, must be terminated with a newline character (`\verb$\n$').
Ignoring such terminators and processing data in blocks with a fixed size of $1,024$ bytes can result in a high probability of processing truncated data, which might lead to incorrect classification of the represented information.
Therefore, identification and transformation must be performed on the data in a format dependent and semantically consistent manner, and \ac{PPFS} achieves this by loading and processing data with a strategy respecting file format specifications and encoding.

For this reason, minimal support for common file formats like \ac{CSV}, XLS, or \acs{JSON} positively impacts the precision of detection and utility preservation, by reducing the risk of incorrect classification of data blocks, and the probability of damaging the file structure.

When a read or write operation is intercepted, \ac{PPFS} inspects the raw data that is being read (or written) in order to identify potentially sensitive information.
On read, \ac{PPFS} retrieves a certain amount of bytes before and after the buffer requested by the client application. We call this additional amount of bytes \emph{guard}.
This allows the identification of sensitive patterns that are expanding beyond the acquired buffer. 
The application of guards is in addition to the ability to support specific file formats, allowing \ac{PPFS} to also handle exotic file formats.
The client application receives only the amount of data initially requested while the other bytes are kept internally by \ac{PPFS} as a cache.
Thus, improving retrieval time for sequential read.

On the other hand, on write, \ac{PPFS} delays the flush operation to be able to perform detection of sensitive information beyond the individual buffer.

These operations require \ac{PPFS} of being aware of all applications accessing files within the directory exposed by \ac{PPFS}, basically mimicking the behaviour of modern operating systems.
The size of the left and right guards can be defined by the user based on empirical observations, or be predefined by the file type.
The latter strategy allows better precision and utility preservation, while requiring the ability to correctly identify file types with extensive work to expand the support for unconventional, or custom file types.

The definition of sensitive information is provided through a \ac{KB}.
The \ac{KB} contains information about the definition of sensitive information, and instructions on how the identified information should be treated.
\ac{PPFS} supports several types of data transformation, ranging from simple redaction, where the  identified values are being replaced with blanks or `\verb|*|', to semantic and format preserving masking~\cite{DBLP:conf/icde/AntonatosBHGA18} and anonymisation techniques such as data generalisation and local differential privacy.

The main advantage of the proposed approach lies in the transparency that the solution provides.
\ac{PPFS} can be deployed as a protection layer in order to reduce the privacy risk in a number of scenarios.
For example, by providing access to data files for monitoring purposes to a third party system, while preventing leakage of incorrectly handled information produced by applications in testing or debug mode.
This way, it is not required to modify the application consuming the data, as \ac{PPFS} can be transparently deployed to inspect, and redact, the data that such an application is consuming and/or producing within a specific portion of file system. 
The sole effort required is to properly define and validate in the \ac{KB} the specifications for detection and transformation of sensitive information.
As an example of such validation, transformations applied need to be consistent with the original data format, if the expected applications are sensitive to data format.

\section{Implementation Details}
\label{sec:implementation}
A prototype has been implemented to validate the feasibility of \ac{PPFS} and to test its impact on performance.

Following agile best practices, we concentrated on creating a proof of concept implementation.
This means leveraging languages and framework that would speed up the development and testing of the system.

For this reason, we created a prototype using \texttt{Python} (version 3.6) and \texttt{python-fuse} as main development library.\footnote{\url{https://github.com/libfuse/python-fuse}}
This library exposes the Python bindings of FUSE~\cite{vangoor2017fuse}. 
The reasons behind these choices are as follows. Python is a popular language for rapid prototyping, thus allowing fast experimentation of various strategies for rules and transformation application.
Similarly, \ac{FUSE} is the de-facto standard for user-space applications exposing a filesystem interface.

These implementation choices have known drawbacks.
Namely, using Python as main language introduces performance penalties, which can be overcome by implementing the application in a more canonical system language (mainly C or C++).
Similarly, the fact that the main functionality of \ac{PPFS} are executed in user space introduces another performance penalty, as we will present and discuss in the evaluation section.
An implementation in a more canonical system language would have yielded better performance, however, we are accounting for this in the evaluations.

The prototype consists of a main application that is in charge of running \ac{FUSE}.
Invoking the \texttt{dlpfs} module from Python requires three mandatory parameters, namely: \texttt{-t}, specifying the file system type; \texttt{-r}, specifying the root directory; \texttt{-m}, specifying the mounting point; and optionally \texttt{-s}, that is the path to the \emph{behaviour specification file}.

Currently, the prototype supports two types of file systems: (i) \texttt{dlpfs}, and (ii) \texttt{\acl{LB}}. The latter is a simple \ac{LB} file system that mirrors the content of the root directory to the mounting directory, and its purpose is only to fairly benchmark \ac{PPFS}, as will be discussed in Section~\ref{sec:evaluation}. The former, \texttt{dlpfs}, is the actual implementation of the method presented in Section~\ref{sec:method}.

The behaviour specification file contains instructions to \ac{PPFS} regarding which data flow to protect (write, read, or both), which patterns to protect and what transformation to apply to the detected patterns.
A simple example of this specification file is presented in Figure~\ref{fig:bsf}.
The structure of the file is simple. It consists of a \ac{JSON}~\footnote{\url{https://www.json.org}} object containing the following fields:
\begin{itemize}
    \item \texttt{do\_read}, a boolean value indicating if read data flow should be protected
    \item \texttt{do\_write}, a boolean value indicating if write data flow should be protected
    \item \texttt{rules}, a list of rules to be applied on read and/or write operations.
\end{itemize}

Each rule is a \ac{JSON} object containing two fields:
\begin{itemize}
    \item \texttt{patterns}, a list of patterns identified within this rule
    \item \texttt{transformation}, the transformation to apply to the detected bytes
\end{itemize}
Currently, \ac{PPFS} supports two types of patterns: regular expressions -- implemented using the Python wrapper for \texttt{re2}\footnote{\url{https://github.com/google/re2/}} -- and lookup tables.
Other types of patterns -- for instance those presented in~\cite{DBLP:conf/icde/AntonatosBHGA18} -- are envisioned to be added to the system according to the needs presented in use cases.

\ac{PPFS} currently supports a small but functional set of transformations: redaction, masking, generalisation, and anonymisation.
Redaction is implemented as a specialisation of masking where the detected bytes are replaced with a predefined character, set as default to `\verb|*|`, preserving the length of the replaced bytes.
Masking, on the other hand, replaces the value with another fictionalised value within the same domain~\cite{DBLP:conf/icde/AntonatosBHGA18,weiss2015practical}.
Generalisation is a special type of masking, where the identified value is replaced with a more generic value within the same domain, for example replacing the value ``Single'' with ``Not Married'', when protecting values within the \emph{Marital Status} domain. Generalisations are performed using external knowledge bases like type hierarchies.
Finally, \ac{PPFS} supports a lightweight form of local differential privacy.
This is achieved by replacing numerical values with the output of the application of a differential privacy mechanism~\cite{DBLP:journals/corr/abs-1907-02444}.



\begin{figure}[tb]
 \centering
\begin{lstlisting}[language=json,numbers=none,basicstyle=\footnotesize]
{  
    "do_read": true,
    "do_write": false,
    "rules": [{
        "patterns": [{
	        "type": "re",
	        "spec": "(:?\\w|\\.)+@(?:\\w|\\.)+\\.\\w{2,4}"
        }],
        "transformation": {
            "type": "redact"
        }}, {
        "patterns":[{
            "type":"re",
            "spec": "Account\\s+total:\\s+(-?\\d+\.\\d{2})"
        }],
        "transformation": {
            "type": "diff_priv",
            "mechanism": "laplace",
            "e": 0.01,
            "d": 0.2
        }
    }]
}
 \end{lstlisting}
 \caption{Example of behaviour specification file content}
 \label{fig:bsf}
\end{figure}

\section{Experimental Evaluation}
\label{sec:evaluation}
This section describes the evaluation setup used to validate the performance of \ac{PPFS}.

\subsection{Setup}
\label{sec:evaluation-setup}

A number of experiments have been conducted in order to assess the impact of \ac{PPFS} on the performance of read and write operations.
The benchmarks presented and discussed in the remainder of this section have been executed on a \ac{VM}, equipped with an Intel\textsuperscript{\textregistered} Xeon\textsuperscript{\textregistered} Gold 6140 CPU \@ $2.30$GHz vCPU with $8$ vcores, $32$GB of RAM, and \ac{SAN} drives.

This scenario mimics a common production environment, where applications are running in a virtualised environment and the hardware stack is abstracted to the user.
It is not uncommon for the storage system of such virtualised environments to be mounted as a remote filesystem, leveraging technologies such as \ac{NFS}.\footnote{\url{https://tools.ietf.org/html/rfc7530}}

Thanks to this approach, for instance, directories can be easily shared across different virtual machines within the same cloud infrastructure, and data can easily migrate across different environments.

This introduces additional penalties to the performance of read and write operations through network factors such as latency, jitter, and congestion.
Therefore, it is paramount to define an unbiased and clear baseline for performing objective and accurate benchmark measurements.

As described in Section \ref{sec:implementation}, the initial \ac{PPFS} prototype has been developed by extending the \texttt{python-fuse} library. The performance impact of the protection offered by \ac{PPFS} has then been measured by comparing its throughput with that obtained from a simple \ac{LB} filesystem implementation that also extends the \texttt{python-fuse} library.

A \emph{\acl{LB}} file system is a simple pseudo-file system implementation that accesses content from a storage device at a given path, and renders it available at a different path.
In other words, it simply forwards read and write operations without introducing any additional computational steps.
For any given benchmark test, the performance of such \ac{LB} implementation has been used as a baseline for the experiments, thus accounting in the comparison for the computational overhead caused by using \texttt{python-fuse} library and network induced delays.

\subsection{Methodology}
\label{sec:evaluation-methodology}

The experiments have been conducted as follows.

First, a number of synthetic datasets have been generated using the Python library \texttt{faker}\footnote{\url{https://github.com/libfuse/python-fuse}}, a popular open source library for the generation of synthetic data.
The data schema of these datasets is the following: 
\begin{itemize}
    \item \texttt{id} contains a monotonically increasing sequence number.
    It reflects the typical row identifier present in most datasets.
    Its values range from $0$ to $N - 1$, where $N$ is the number of rows contained in the dataset.
    \item \texttt{icd} contains a valid \ac{ICD} value with probability $0.05$, or an empty string.
    The \ac{ICD} is an international coding standard maintained by the \ac{WHO}, which is globally used as diagnostic standard for epidemiology, health management, and clinical purposes. This field contains valid values for the version $10$ of the standard.\footnote{\url{https://icd.who.int/browse10/2019/en}}
    \item \texttt{amount} contains a randomly generated currency value.
    Its values range from $1$ to $1,000$ US dollars, with up to $2$ decimal places.
    The values are sampled uniformly from the domain.
    \item \texttt{message} contains a variable length string representing a text message, or a comment, and it is composed by concatenating:
        \begin{itemize}
            \item A randomly generated sentence, with length varying between $3$ and $9$ words.
            \item A first keyword with probability $0.01$.
            \item A second keyword with probability $0.1$.
            \item A randomly generated email address with probability $0.05$.
            \item Another randomly generated sentence, 
            comprised of $3$ to $9$ words.
        \end{itemize}
\end{itemize}

The test data is then represented in \ac{CSV} format \footnote{\url{https://tools.ietf.org/html/rfc4180}} and an excerpt of a test dataset is shown in Figure~\ref{fig:test-data}.
Note that rows are truncated with \texttt{(...)} for readability purposes.

\begin{figure*}
\centering
\begin{tabular}{c}
\begin{lstlisting}[basicstyle=\small,frame = single,linewidth=0.8\textwidth]
...
124,"G30.1","$683.91","Force food second. Direction note his finish case."
125,"C00.6","$3.97","Carry wish quickly industry... International visit..."
126,"F71.8","$355.56","The politics mother resource... Charge fill that..."
127,"D51.3","$93.64","Born industry here... Health ever nearly achieved..."
128,"G29.3","$87.94","Role method must... FrequentKeyword. Late why hold..."
129,"F71.1","$159.71","Father go everybody... Big according he move."
130,"B20.3","$874.19","Chance data under line left... FrequentKeyword..."
131,"C00.2","$825.05","Nation cut last old... vanessa36@cox-mata.net..."
...
\end{lstlisting}
\end{tabular}
 \caption{Example of generated data.}
 \label{fig:test-data}
\end{figure*}


We created several of such \ac{CSV} test datasets with sizes ranging from $1$ to $20,000$ rows, where each row amounts to approximately $100$ bytes, and we then performed two main batches of experiments.

The first batch concentrates on exploring the performance impact of \ac{PPFS} on read operations, while the second one concentrates on measuring the impact on write operations.

\paragraph{Read Strategies}
\label{sec:read-strategies}

We tested a number of read strategies, with the objective of simulating behaviours that are commonly followed by applications while reading the content of an input file.
Namely, we simulated the following scenarios:
\begin{itemize}
    \item Entire file content loaded in memory as \texttt{pandas}\footnote{\url{https://pandas.pydata.org/}} dataframe, this strategy replicates the usual behaviour of a data scientist or machine learning practitioner.
    \item Entire file content entirely copied in memory, another common practice to load and process the content of files
    \item Scan file content one row at a time, delegating to Python the identification of row boundaries, typically via the new line character (\verb|\n|). This is the behaviour of row-oriented programs or scripts.
    \item Read file content using \ac{OS} instructions with varying read buffer size between $10$, $100$, $1,000$, and $10,000$ bytes.
    This strategy simulates sequential access to file when loading fixed size buffers, for example when data objects are deserialised from disk.
\end{itemize}

\paragraph{Write Strategies}
\label{sec:write-strategies}

Similarly to how the read performances were tested, we also executed benchmarks of different behaviours with respect to writing files 
to disk.
Namely, we simulated the following writing patterns:
\begin{itemize}
    \item Entire file content written to disk as \texttt{pandas} dataframe, this strategy replicates the usual behaviour of a data scientist or machine learning practitioner who is storing the result of a computation to disk.
    \item Entire file at once, this pattern simulates an application saving the all the output at once, or a program faulting and creating a memory dump.
    \item Row by row, this pattern mimics the behaviour of an application periodically logging messages to disc.
    \item Field by field, this pattern replicates the behaviour of an application incrementally writing the produced output.
\end{itemize}

\paragraph{\ac{PPFS} configuration}
As we will show later in this section, the most important factor on the performance of \ac{PPFS} resides in its configurations, in terms of identification pattern and guard sizes.

We tested numerous configurations.
First of all, we tested the overhead caused by the \ac{PPFS} architecture. This has been done providing a configuration with no patterns or transformation.
After that we tested with different types of patterns, namely regular expressions of various complexity and coverage.
We tested the impact of an administrator specifying not optimised regular expressions (i.e. containing unnecessary greedy operators, or containing overlapping parts) against precise patterns.

We then tested the performance impact of using different guard sizes, ranging from $0$ (i.e. no guard) to $256$ bytes.

\paragraph{Matching cases}
\label{sec:matching}

The last variable in our evaluation is the percentage of matches encountered by the privacy protection policies when executing the read or write operation.
More precisely, we tested several policies that differ in the number of matching patterns with the file that is being read or written.
As it will be presented in Section~\ref{sec:results}, this is one of the factors that most impacted performance of the system.
We tested three main cases:
\begin{itemize}
    \item \textbf{No matches.} Thus, specifying patterns that were by design not existing in the test data.
    \item \textbf{Few matches.} In this case we used a set of patterns having a low probability of match within the test data. More precisely, we tested patterns with probability $0.01$ of being present in the test data, according to both data construction and post data generation assessment.
    \item \textbf{Many matches.} In this case we used a set of patterns having a higher probability of match within the test data. More precisely, we tested patterns with probability $0.10$ of being present in the test data, according to both data construction and post data generation assessment.
\end{itemize}

Finally, we also tested different strategies in terms of how the patterns are matched, and how the patterns behave.
For example, we noticed in the preliminary evaluation how the structure of patterns implemented as regular expression produced very different results depending on whether the regular expression itself had certain characteristics.
As one would expect, optimised regular expressions with less overlapping parts and less greedy operators were performing better.

\subsection{Results and Discussion}
\label{sec:results}

\begin{figure*}[hp]
    \centering
    \subfloat[Read strategy]{\includegraphics[width=.45\textwidth]{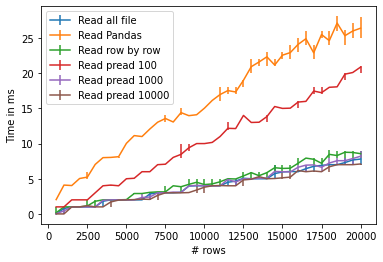}}
    \quad
    \subfloat[Write]{\includegraphics[width=.45\textwidth]{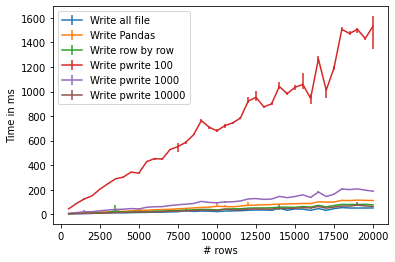}}
    \caption{\ac{LB} performance, varying file size, all strategies}
    \label{fig:base-file-size}
\end{figure*}
\begin{figure*}[hp]
    \centering
    \subfloat[Read: No pattern specified\label{fig:overhead:norules}]{\includegraphics[width=.41\textwidth]{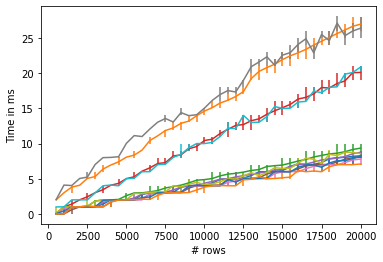}}%
    \subfloat[Read: No matching pattern\label{fig:overhead:nomatching}]{\includegraphics[width=.55\textwidth]{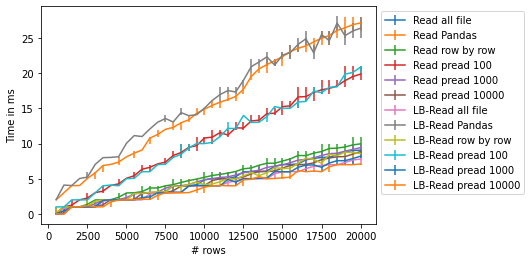}}\\
    \subfloat[Write: No pattern specified\label{fig:overhead-write:norules}]{\includegraphics[width=.41\textwidth]{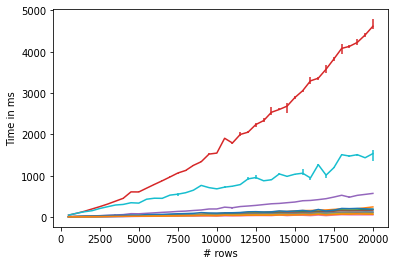}}
    \subfloat[Write: No matching pattern\label{fig:overhead-write:nomatching}]{\includegraphics[width=.55\textwidth]{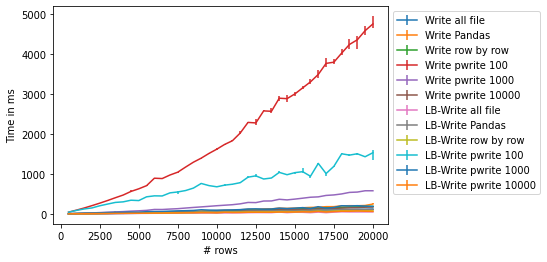}}
    \caption{Minimum penalty of \ac{PPFS} over \ac{LB}}
    \label{fig:overhead}
\end{figure*}
\begin{figure*}[hp]
    \centering
    \subfloat[Not optimised\label{fig:read-size-cat:all}]{\includegraphics[width=.45\textwidth]{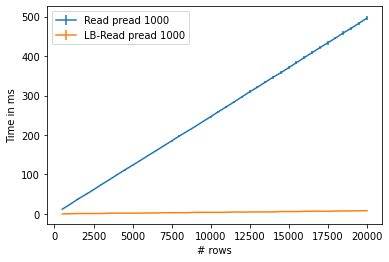}}
    \quad
    \subfloat[Many match\label{fig:read-size-cat:many}]{\includegraphics[width=.45\textwidth]{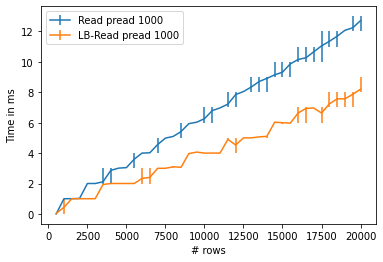}}\\
    \subfloat[Few match\label{fig:read-size-cat:few}]{\includegraphics[width=.45\textwidth]{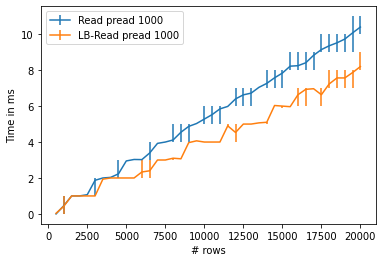}}
    \quad
    \subfloat[Dictionary\label{fig:read-size-cat:dictionary}]{\includegraphics[width=.45\textwidth]{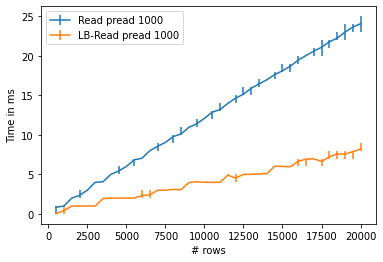}}
    \caption{Read performance, varying file size, pread-1000}
    \label{fig:read-size-cat}
\end{figure*}
\begin{figure*}[hp]
    \centering
    \subfloat[Many matches\label{fig:read-guard-various:many}]{\includegraphics[width=.45\textwidth]{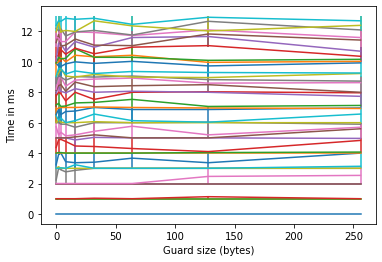}}
    \quad
    \subfloat[Few match\label{fig:read-guard-various:few}]{\includegraphics[width=.45\textwidth]{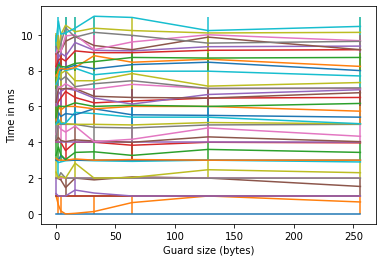}}
    \caption{Read performance, varying guard size}
    \label{fig:read-guard-various}
\end{figure*}
\begin{figure*}[hp]
    \centering
    \subfloat[Not optimised\label{fig:write-size-cat:all}]{\includegraphics[width=.45\textwidth]{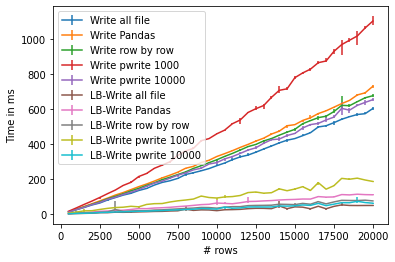}}
    \quad
    \subfloat[Many match\label{fig:write-size-cat:many}]{\includegraphics[width=.45\textwidth]{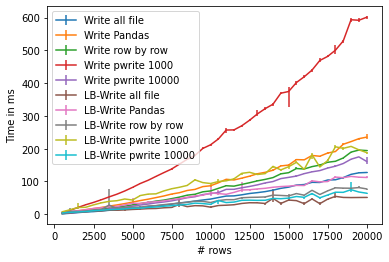}}\\
    \subfloat[Few match\label{fig:write-size-cat:few}]{\includegraphics[width=.45\textwidth]{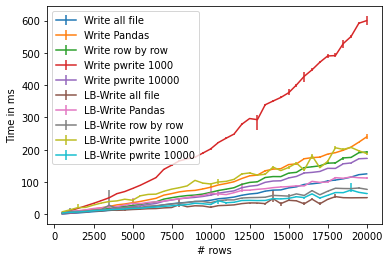}}
    \quad
    \subfloat[Dictionary\label{fig:write-size-cat:dictionary}]{\includegraphics[width=.45\textwidth]{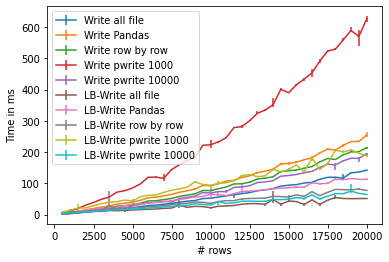}}
    \caption{Write performance, varying file size
    }
    \label{fig:write-size}
\end{figure*}
\begin{figure*}[hp]
    \centering
    \subfloat[Read pattern\label{fig:timeline:read}]{\includegraphics[width=.45\textwidth]{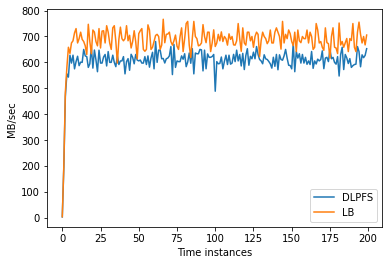}}
    \quad
    \subfloat[Write pattern\label{fig:timeline:write}]{\includegraphics[width=.45\textwidth]{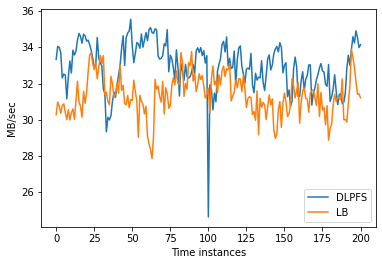}}
    \caption{Application throughput}
    \label{fig:timeline}
\end{figure*}

We repeated executions of the bulk of experiments $30$ times, and report the mean, 10th and 90th percentile of the measured results.

Before analyzing the performance of \ac{PPFS}, let us argue about more general observation.
First of all, the experimental evaluation clearly shows the importance of the correct selection of the detection engine.
One might notice how, with no detection, \ac{PPFS} behaves exactly as the \ac{LB} baseline, which means that the additional buffering is not impacting overall read/write performances. 
Let us also remark that the time taken by the actual transformation is negligible when compared with the detection, as demonstrated by preliminary execution of \ac{PPFS} with configuration specifying no transformation, redact, masking (randomisation) or differential privacy noise addition.
Table~\ref{table:transform} presents the average of $30$ runs over randomly generated $20,000$ numerical values transformed with the strategies supported by \ac{PPFS}. 
In fact, the average time required for processing $20,000$ numerical values takes is, $3.459$, $6.113$, $151.982$, and $319.207$ milliseconds.
The only exception is the application of noise addition in a differentially private fashion. This is caused by two factors. The first one relates to the fact that the used framework has been designed to operate on vectors of values, not individual ones. Secondly, the framework is designed to sample noise from a distribution in a secure manner~\cite{Mir12,holohan2021secure}, a procedure that introduces additional complexity.

\begin{table}[hp]
    \centering
    \caption{Transformations time.}
    \label{table:transform}
    \begin{tabular}{| c | c |}
         \hline
         \textbf{Transformation} & \textbf{Time (ms)} \\
         \hline\hline
         No transformation & 3.4593\\
         Redact & 6.11317\\
         Masking & 151.982\\
         DP noise & 319.207\\
         \hline
    \end{tabular}
\end{table}

On the other hand, as patterns are detected the reader should notice an increment in execution time.
The amount of execution time directly depends on the amount of matches the pattern has in the file, as one would expect.
Moreover, the engine actually used for the detection of the patterns greatly impacts the amount of time spent in this phase.
This analysis of difference in performance, for example between \texttt{re} and \texttt{re2},
is beyond the scope of this paper and it has been previously discussed.\footnote{\url{https://pypi.org/project/re2/\#performance}}
For the rest of this we will present only the best performing detection engine.

Similarly, the strategy of operation affects the performance.
For example, Figure~\ref{fig:base-file-size} shows the difference in time required to process different files reading, or writing, using the different strategies.
A common pattern that can be observed, is that the time increases linearly as the file sizes increase.
This is an expected behaviour, and follows the trend of the baseline, even if generally with a steeper slope.
This is shown in Figure~\ref{fig:overhead}, where we present the trend for two special cases.
The first one, Figure~\ref{fig:overhead:norules}, where the protection policy is set to empty, and a second one, Figure~\ref{fig:overhead:nomatching}, where the policy has no match in the processed data.
The experiments performed on the write path provide a similar picture, although the overhead of validating data on write is greater than for read, as shown in Figure~\ref{fig:overhead-write:norules} and Figure~\ref{fig:overhead-write:nomatching}.

Figure~\ref{fig:read-size-cat} shows the execution time of the introduced policies when the entire file is read as a block of data.
The first observation is that the guard size does not seem to impact significantly the performance.
On the other hand, the specified patterns greatly impact the overall performance, as clearly shown in Figure~\ref{fig:read-size-cat} and following.
A poorly optimised set of patterns, as shown in Figure~\ref{fig:read-size-cat:all}, reduces the system performance greatly, while a set of patterns with similar hit ratio but with more optimised regular expressions still shows a significant but from a practical viewpoint acceptable overhead (see Figure~\ref{fig:read-size-cat:many}).
On the other hand, in case of patterns with few hits in the data, the performances are affected by less then $30\%$, as shown in Figure~\ref{fig:read-size-cat:few} and Figure~\ref{fig:read-size-cat:dictionary}.

Figure~\ref{fig:read-guard-various} shows how the performance changes with the guard size.
Once can notice how there is no significant variation as the guard, which we remind is the amount of bytes \ac{PPFS} reads before and/or after the buffer required by the user, ranges from 0 to 256 bytes. One might only notice a shift on the y-axis caused by the different number of matches between Figure~\ref{fig:read-guard-various:many}, having many matches, and Figure~\ref{fig:read-guard-various:few}, having fewer matches.

Similar as for read, also the write pattern performances are mostly influenced by the pattern itself and the privacy protection policy enforced.
Figure~\ref{fig:write-size} presents an overview of the impact.
The performance can degrade up to twice in case of many matches, as shown in Figure~\ref{fig:write-size-cat:many}, but can be deemed generally acceptable for non real-time services.

After this analysis, we can conclude the \ac{PPFS} has known costs in terms of performance, but compares favourably considering the additional protection provided.
In fact, the mentioned motivating scenario assumes \ac{PPFS} to be deployed as additional protection layer, thus generally providing a minimal impact on application's performance as shown in Figure~\ref{fig:overhead}, while providing additional guarantees in rare but critical events.
This is further corroborated by Figure~\ref{fig:timeline}, where we present the throughput of an application reading (Figure~\ref{fig:timeline:read}) and writing (Figure~\ref{fig:timeline:write}) data.
This application behaves according to the following pattern:
first it reads(/writes) non sensitive data. At $t=100$ the application accesses a protected pattern, after which it resumes normal operation.

\section{Related Work}
\label{sec:related}
Properly protecting data outsourcing or sharing, even locally, is an open issue.
Several works have been proposed to address these issues in specific context, with particular focus on context where sensitive data are pervasive, like in the healthcare domain~\cite{appari2010information}.

The majority of the proposed approaches leverage, one way or another, cryptographic-based techniques.
For example, ~\cite{wang2016protecting} presents a cryptographic-based access control mechanism to selectively limit access to sensitive parts of the file.
Similarly, \cite{sicuranza2013secure,ciampi2013federated} describe a system, and associated architecture, to introduce cryptography-based techniques in federated health information systems. The authors show the feasibility of improving the security of such systems by adopting proper mechanisms to protect the exchanged data and the provided functionalities from malicious manipulations.
Still in the healthcare domain, other approaches -- like the one presented in~\cite{ijbs24617} -- tackle the problem of data sharing using a microservice approach.
Hence, data is provided on demand using highly restricted access control rules,  to reveal data on a need-to-know bases, and transforming the data in an abstract data format before release, thus limiting the risk of data leakage.

Other approaches rely on different ways to encode the files on storage.
For example, \cite{sheng2011privacy} presents a new file system that focuses on the privacy protection of the on-disk state. 
This is achieved by re-ordering data in user files at the bit level, and storing bit slices at distributed locations in the storage system.
On the other hand, \cite{djoko2019nexus} presents a stackable filesystem that leverages trusted hardware to provide confidentiality and integrity for user files stored on untrusted platforms.
A similar idea is presented in~\cite{paul2015possible}, where the authors propose a technique that involves using a hash function that uniquely identifies the data and then splitting data across multiple cloud providers. This is done following a ``Good Enough" approach to privacy-preserving cloud data storage, which has been proven to be both technologically feasible and financially advantageous.

Moreover,~\cite{7345372} presents a statistical \ac{DLP} model to classify data on the basis of semantics. This study contributes by using data statistical analysis to detect evolved confidential data.
A fairly a summary and comparison of \ac{DLP} systems, techniques and research directions is also provided in~\cite{alneyadi2016survey}.

The work the most similar to \ac{PPFS} is presented in~\cite{hyppocratic}. The authors analyse and propose mechanisms to enhance the disclosure control of personal data. The scheme, called the Hippocratic Filesystem, stores personal data's purpose and use limitation as the data's label, propagates the label as the information flows from one place to another, and enforces the label to prevent accidental disclosures. \ac{PPFS}, on the other hand, presents a complementary method, where data is transformed either at reading or writing time.
Similarly,~\cite{agrawal2002hippocratic} proposes the so called \ac{HDB}, which presents similar concept to the filesystem approach previously presented, but in the context of a centralised database.

Moreover,~\cite{zhang2009research} presents a Windows file system that transparently encrypts files automatically according to encryption strategies. This work is complementary to the approach here presented. The main differentiation is that in \ac{PPFS} it is not mandatory to access the protected data through \ac{PPFS} itself. A file directory can be protected while accessed from some applications, while others can access the data without interacting with \ac{PPFS}, thus introducing a performance penalty only when deemed necessary. 

Leveraging \ac{HDB}, a P2P-based solution to tackle the private data sharing problem in social networks has been presented~\cite{jawad2013supporting}.
The identification and transformation capabilities of \ac{PPFS} are inspired by the work in~\cite{DBLP:conf/icde/AntonatosBHGA18}, which presents a toolkit that contains functionality for the detection and format preserving transformation of values.

Finally, an extensive survey of masking anonymisation and cryptographic-based methods for outsourced data storage is presented in~\cite{domingo2019privacy}. This survey was instrumental to the design of \ac{PPFS} because, even if the application scenario is different, the referenced techniques can be ported to the approach here presented.

\section{Conclusions and Future Work}
\label{sec:conclusion}
We presented \ac{PPFS}, a novel data leakage prevention file system middleware, to protect sensitive information potentially stored in shared systems.
We demonstrated the technical feasibility and experimentally evaluated the performance impact of the system.
In particular, the evaluation demonstrated that little to none overhead is introduced by \ac{PPFS} on normal file-based operations, with reductions in performance detected only when sensitive data is protected.

Future work can focus on four main aspects.
First, scaling up the concepts illustrated here in a purely distributed setting, for example by porting the prototype to Java to enhance \acs{HDFS}.
Second, the extension of the capabilities offered by \ac{PPFS} in terms of data transformation. This could materialise as an integration 
with more established data privacy frameworks.
Third, to extend data detection capabilities, for example with the integration of contextual information, such as file metadata, application and user operation, during the detection process.
Fourth and finally, we envision to integrate \ac{PPFS} with 
conventional access control frameworks, to simplify configuration management and deployment. 


\bibliographystyle{abbrv}
\bibliography{bibliography}

\end{document}